\documentclass[prl,aps,twocolumn,superscriptaddress]{revtex4-1}

\RequirePackage{latexPreamblefp}
\usepackage{graphicx, color}
\begin{document}

\begin{abstract}
It is well-known that in certain scenarios weakly entangled states can generate stronger nonlocal effects than their maximally entangled counterparts.  In this paper, we consider violations of the CHSH Inequality when one party has inefficient detectors, a scenario known as an asymmetric Bell experiment.  
For any fixed detection efficiency, we derive a simple upper bound on the entanglement needed to violate the inequality by more than some specified amount $\kappa\geq 0$.  When $\kappa=0$, the amount of entanglement in all states violating the inequality goes to zero as the detection efficiency approaches $50\%$ from above.  
We finally consider the scenario in which detection inefficiency arises for only one choice of local measurement.  In this case, it is shown that the CHSH Inequality can always be violated for any nonzero detection efficiency and any choice of non-commuting measurements.

\end{abstract}

\author{Daniel Dilley}\email{dilleyd@siu.edu}
\affiliation{Department of Physics and Astronomy, Southern Illinois University,
Carbondale, Illinois 62901, USA} 
\title{More nonlocality with less entanglement in Clauser-Horne-Shimony-Holt experiments using inefficient detectors}   
\author{Eric Chitambar}\email{echitamb@siu.edu}
\affiliation{Department of Physics and Astronomy, Southern Illinois University,
Carbondale, Illinois 62901, USA}

\date{\today}

\maketitle

Nonlocality and entanglement can be seen as two distinct features of multi-part quantum systems.  Building upon seminal work conducted by Werner \cite{Werner-1989a}, certain entangled states have been shown to generate local measurement statistics that can be simulated by a local hidden variable (LHV) model  \cite{Barrett-2002a, Bowles-2016a}.  These entangled states cannot violate a Bell Inequality, and thus the subsystems fail to share nonlocal correlations \cite{Bell-1964a}.

Beyond this fundamental discovery, the precise relationship between entanglement and nonlocality is not adequately understood.  Intuitively, one would expect that states become more nonlocal as they become more entangled.  However, there are two reasons why this intuition fails.  First, in general there is no unique measure for either entanglement or nonlocality \cite{Horodecki-2009a}, and so there is no unambiguous way to claim that one state is more entangled or more nonlocal than another.  Second and more importantly, even after fixing a natural choice of measures, certain nonlocal effects are indeed more prominent in weaker entangled states compared to their stronger counterparts \cite{Eberhard-1993a, Hardy-1993a, Acin-2002a, Brunner-2005a, Cabello-2007a, Brunner-2007a, Pal-2010a, Vidick-2011a, Junge-2011a, Liang-2011a, Vallone-2014a, Palazuelos-2014a, Christensen-2015a}.  M\'{e}thot and Scarini have referred to this phenomenon as an ``anomaly of nonlocality'' \cite{Methot-2007a} (or perhaps one prefers to call this an ``anomaly of entanglement'').  

The earliest evidence for this anomaly came from work by Eberhard who studied the violation of the Clauser-Horne-Shimony-Holt (CHSH) Inequality \cite{Clauser-1969a} in the presence of detector inefficiencies \cite{Eberhard-1993a}.  Inefficient detectors in CHSH experiments allow for LHV theories to exploit the so-called ``detection loophole'' and spuriously simulate nonlocal correlations \cite{Pearle-1970a}. Through numerical optimization, Eberhard showed that as all detectors in a CHSH experiment become more inefficient, weaker entangled states are needed to violate the CHSH Inequality.  This was later confirmed analytically by Liang \textit{et. al} \cite{Liang-2011a}.  Thus if robustness to imperfect measurement is regarded as a nonlocality measure, nonlocality and entanglement are inversely related in the CHSH setting (see also \cite{Vallone-2014a}).

In this paper we derive a transparent trade-off between the amount of entanglement in a state and the degree to which it can violate the CHSH Inequality with detector inefficiency.  This is accomplished by considering a hybrid scenario in which Alice (who holds one subsystem) has perfect detection efficiency, but Bob (who holds the other) has a conclusive detection only a fraction $\eta$ of the time.  This is often described as an asymmetric Bell experiment, and it has been previously proposed as a useful model to study the entanglement in atom-photon systems \cite{Cabello-2007a, Brunner-2007a}.  Unlike the difficulty in attaining high detection of single photons, atomic measurements can be made with efficiency close to one.  Thus with all other considerations besides detection efficiency being equal, circumventing the detection loophole in CHSH experiments becomes easier using atom-photonic entanglement.  

For asymmetric Bell tests, it was shown that the CHSH Inequality can always be violated for any one-sided efficiency $\eta>1/2$ \cite{Cabello-2007a, Brunner-2007a}.  In this paper, we derive a simple analytic upper bound on the amount of entanglement needed to violate the CHSH Inequality by a certain amount for a given $\eta$.  This upper bound goes to zero as $\eta\to 1/2$, and thus smaller amounts of entanglement are necessary and sufficient to demonstrate nonlocality as the detection efficiency decreases.  We further consider the scenario where Bob has detection inefficiency for only one choice of measurement.  Surprisingly, for any $\eta>0$, the CHSH Inequality can be violated by some quantum state using \textit{any} pairs of non-commuting local observables for Alice and Bob.  

In a traditional CHSH experiment, multiple copies of a two-qubit state $\rho$ are generated and distributed to spatially separated Alice and Bob.  On each copy, Alice randomly chooses to measure the spin of her qubit in either direction $\hat{a}_0$ or $\hat{a}_1$, and likewise Bob randomly measures in either direction $\hat{b}_0$ or $\hat{b}_1$.  Under ideal conditions, each measurement always generates a single spin-up or a spin-down outcome, denoted by ``$0$'' and ``$1$'' respectively.  Then if Alice and Bob share quantum state $\rho^{AB}$, the conditional probabilities of outcomes $a,b\in\{0,1\}$ for measurement choices $x,y\in\{0,1\}$ are given by
\begin{align}
\hat{p}(a,b|x,y)&=\tr\left[\rho^{AB}(\Pi^A_{a|x}\otimes \Pi^B_{b|y})\right],
\end{align}
where 
\begin{align}
\Pi^A_{a|x}&=\frac{1}{2}(\mbb{I}+(-1)^a \hat{a}_x\cdot\hat{\sigma})\label{Eq:Proj-Alice}
\end{align}
\begin{align} 
\Pi^B_{b|y}&=\frac{1}{2}(\mbb{I}+(-1)^b \hat{b}_y\cdot\hat{\sigma})\label{Eq:Proj-Bob}
\end{align}
 are projectors and $\hat{\sigma}=\sigma_x\hat{x}+\sigma_y\hat{y}+\sigma_z\hat{z}$ is a vector of Pauli matrices.  

However, in most realistic setups, Alice or Bob may not obtain a conclusive measurement outcome for each copy of $\rho^{AB}$ and for each choice of measurement; this is known as detection inefficiency, and an inconclusive outcome can be denoted by $\emptyset$.  There are two standard ways to deal with detection inefficiencies \cite{Brunner-2014a}.  One could simply discard all inconclusive measurement outcomes and compute the measurement statistics using the events when both Alice and Bob obtain conclusive outcomes. This process is known as post-selection, and its ability to certify genuine nonlocal correlations requires the additional assumption, known as ``fair sampling,'' that the post-selected events faithfully represent the ideal scenario of perfect detection efficiency \cite{Pearle-1970a, Clauser-1974a}.  

A second approach does not use post-selection and simply involves coarse-graining over the inconclusive outcome $\emptyset$ and one of the other conclusive outcomes $\{0,1\}$. The result is a non-projective positive-valued operator measure (POVM) with two outcomes.  In this paper, we consider the partially idealized scenario where an inconclusive outcome can arise only on Bob's measurements.  First we analyze the case where an inclusive outcome occurs with frequency $(1-\eta)$ for both of his measurement choices.  When combining the inconclusive outcome with the $b=0$ outcome, Bob's POVMs in Eq. \eqref{Eq:Proj-Bob} are replaced with the POVMs $\{\wt{\Pi}^B_{0|y},\wt{\Pi}^{B}_{1|y}\}$, where
\begin{align}
\wt{\Pi}^B_{0|y} &=  \frac{\eta}{2}\left(\mbb{I}+\hat{b}_y\cdot\hat{\sigma}\right)+\left(1-\eta\right)\bbI,\label{Eq:Bob-POVM1}\\
\wt{\Pi}^B_{1|y} &= \frac{\eta}{2}\left(\mbb{I}-\hat{b}_y\cdot\hat{\sigma}\right)\label{Eq:Bob-POVM2}
\end{align}
for $y\in\{0,1\}$.  Note that the term $(1-\eta)\mbb{I}$ in $\wt{\Pi}^B_{0|y}$ corresponds to the inconclusive outcome. 

The original CHSH experiment with detection efficiency $\eta$ on Bob's side can thus be described as effectively an ideal CHSH experiment where Alice chooses between two measurements each having the form of Eq. \eqref{Eq:Proj-Alice}, and Bob chooses between two measurements each having the form of Eqns. \eqref{Eq:Bob-POVM1}-\eqref{Eq:Bob-POVM2}.  In what follows, it will be convenient to introduce the observables $O^A_x=\Pi^A_{0|x}-\Pi^A_{1|x}$ and $O^B_y=\wt{\Pi}^B_{0|y}-\wt{\Pi}^B_{1|y}$.  Then for local measurement choices $(x,y)$ on state $\rho^{AB}$, the expected value of the nonlocal function $f(a,b)=(-1)^{ab}$ computed from the measurement outcomes is 
\begin{align}
\notag
&E(x,y): \; =\sum_{a,b=0}^1f(a,b)p(a,b|x,y)=\tr\left[\rho^{AB}O^A_{x}\otimes O^B_y\right]. \\ \notag \\ \notag
&\text{The CHSH Inequality \citep{Clauser-1969a} says that} 
\end{align}
\begin{align}
\label{Eq:CHSH}
2 &\geq \left|E(0,0)+E(0,1)+E(1,0)-E(1,1)\right| \\
&=\left|\tr[\rho^{AB}\mc{B}]\right|,
\end{align} 
where 
\[\mc{B}=O^A_0\otimes(O^B_0+O^B_1)+O^A_1\otimes(O^B_0-O^B_1)\]
 is the Bell operator for the measurement directions $\{\hat{a}_0,\hat{a}_1,\hat{b}_0,\hat{b}_1\}$ and detection efficiency $\eta$ on Bob's side.  When $\eta=1$, we recover the standard Bell operator \cite{Cirel'son-1980a}, and we denote this by $\mc{B}_{CHSH}$.

Our ultimate interest lies in understanding the entanglement of states that violate Eq. \eqref{Eq:CHSH}.  Since local unitaries (LU) do not change the entanglement in a quantum state, it can be assumed without loss of generality that $\hat{a}_0=\hat{z}$, $\hat{a}_1=\cos\theta_A\hat{x}+\sin\theta_A\hat{z}$, $\hat{b}_0=\hat{z}$, and $\hat{b}_1=\cos\theta_B\hat{x}+\sin\theta_B\hat{z}$ for some angles $\theta_A$ and $\theta_B$.  This is a simple consequence of the fact that Alice and Bob's local Bloch spheres can be arbitrarily rotated using local unitaries.  Thus, the Bell operator reduces to the form
\begin{align}
\label{Eq:Bell-Op1}
\mc{B}&=\sum_{i,j\in\{x,z\}}c_{ij}\sigma_i\otimes\sigma_j+r_z\sigma_z\otimes\mbb{I}\notag\\
&=\eta\mc{B}_{CHSH}+2(1-\eta)\sigma_z\otimes\mbb{I},
\end{align}
where
\begin{align}
r_z&=2(1 - \eta)\hspace{24mm} c_{xx}=-\eta \cos\theta_A\cos\theta_B\notag\\
c_{xz}&=\eta\cos\theta_A(1-\sin\theta_B)\qquad c_{zx}=\eta \cos\theta_B(1-\sin\theta_A)\notag\\
c_{zz}&=\eta(1+\sin\theta_A+\sin\theta_B-\sin\theta_A\sin\theta_B).\notag
\end{align}
Notice that $(\sigma_y\otimes\mbb{I})\mc{B}(\sigma_y\otimes\mbb{I})=-\mc{B}$, which means that a state $\rho$ will violate the CHSH Inequality iff there exists another state $\rho'$ with the same entanglement as $\rho$ and satisfying $\tr[\rho\mc{B}]>2$.  

Let us first show that $\mc{B}$ leads to a violation of the CHSH Inequality whenever $\eta>1/2$.
Suppose $\eta=1/2+\kappa$ for $\kappa>0$, and consider the measurement directions given by $\theta_A=0$ and $\theta_B=\arcsin\frac{(1-2\kappa)^2}{(1+2\kappa)^2}$.   Then, a straightforward calculation shows that the largest eigenvalue of $\mc{B}$ is given by
\begin{align}
\lambda_{\max}=2\sqrt{1+4\kappa^2}.
\end{align}
Hence, $\bra{\lambda_{\max}}\mc{B}\ket{\lambda_{\max}}>2$ whenever $\kappa>0$, where $\ket{\lambda_{\max}}$ is an eigenvalue associated with $\lambda_{\max}$.  

We next analyze the entanglement in states satisfying $\tr[\rho\mc{B}]>2$.  In particular, we are interested in bounding the entanglement of such states as a function of $\eta$.  For convenience, we use the concurrence entanglement measure \cite{Wootters-1998a}, but the same effect can be seen for any other choice of measure.  Let us first focus on pure states.  The concurrence of a given two-qubit state $\ket{\phi}$ is $C(\op{\phi}{\phi})=2\sqrt{\kappa_{\max}(1-\kappa_{\max})}$, where $\kappa_{\max}$ is the largest squared Schmidt coefficient of $\ket{\phi}$.  Note that $\kappa_{\max}$ ranges between $1/2$ and $1$, and the concurrence varies inversely between $1$ and $0$.  One key inequality used in our analysis is a result of Verstraete and Wolf who showed that $\bra{\psi}\mc{B}_{CHSH}\ket{\psi}\leq 2\sqrt{1+C^2}$ \cite{Verstraete-2002a}, where for convenience here and henceforth we let $C$ denote the concurrence of $\ket{\phi}$.  Thus from Eq. \eqref{Eq:Bell-Op1} we have
\begin{equation}
\bra{\phi}\mc{B}\ket{\phi}\leq 2\eta \sqrt{1+C^2}+2(1-\eta)\bra{\phi}\sigma_z\otimes\mbb{I}\ket{\phi}.
\end{equation}
To bound the last term, we write $\ket{\phi} = R \otimes \bbI \ket{\Phi}$, where $\ket{\Phi} = (\ket{00}+\ket{11})$ and $\tr(R R^\dagger) = 1$.  Note that the eigenvalues of $RR^\dagger$ are precisely the squared Schmidt coefficients of the state $\ket{\phi}$.  Then 
\begin{equation}
\label{Eq:noncorrelation_Trace}
\bra{\phi}\sigma_z\otimes\mbb{I}\ket{\phi}=\tr(R R^\dagger \sigma_z )\leq 2 \kappa_{\max}  - 1,
\end{equation}
which follows from the general trace inequality $\tr(A B) \leq \sum_{i = 1}^n \lambda_i (A) \lambda_i (B)$, where $A$ and $B$ are two $n \times n$ hermitian matrices and the $\lambda_i (\cdot)$ are their eigenvalues in non-increasing order \cite{Lasserre-1995a}.  Thus putting together the two bounds we  obtain
\begin{align}
\bra{\psi}\mc{B}\ket{\psi}\leq 2\eta\sqrt{1+C^2}+2(1-\eta)\sqrt{1-C^2}.
\end{align}
Suppose now that $\bra{\psi}\mc{B}\ket{\psi}>2(1+\kappa)$ for some $\kappa\geq 0$.  Then 
\begin{equation}
1+\kappa< \eta\sqrt{1+C^2}+(1-\eta)\sqrt{1-C^2}.\notag
\end{equation}
Squaring both sides gives $(1+\kappa)^2-\eta^2(1+C^2)-(1-\eta)^2(1-C^2) < 2\eta(1-\eta)\sqrt{1-C^4}$.  The LHS is non-negative for all $C$ provided $1/2\leq\eta\leq (1+\kappa)/\sqrt{2}$.  Hence we can square both sides again and after some algebra arrive at the inequality $0 > \alpha C^4 +\beta C^2+\gamma$, with $\alpha=(1-2\eta+2\eta^2)^2$, $\beta=-2(2\eta-1)(\kappa(2+\kappa)+2\eta(1-\eta))$, and $\gamma=\kappa(2+\kappa)(\kappa(2+\kappa)+4\eta(1-\eta))$.  Solving this yields the explicit bound for $C^2$ given in Eq. \eqref{Eq:main_bound}, but we first continue with the solution when $\kappa=0$.  This corresponds to any violation of the locality bound, regardless of the magnitude.  Taking $\kappa=0$ immediately yields the bound
\begin{equation}
\label{Eq:violation_ub}
C^2<\frac{4\eta(1-\eta)(2\eta-1)}{(1-2\eta+2\eta^2)^2}.
\end{equation}
To compute the limit as $\eta\to 1/2^+$, we apply L'H\^{o}pital's rule twice.  Doing so yields the desired result:
\begin{equation}
\lim_{\eta\to 1/2^+}\frac{4\eta(1-\eta)(2\eta-1)}{(1-2\eta+2\eta^2)^2}=0.
\end{equation}
Turning to the case of mixed states, we recall that every two-qubit density matrix $\rho$ has a pure-state decomposition $\rho=\sum_{i}p_i\op{\phi_i}{\phi_i}$ such that $C(\rho)=C(\op{\phi_i}{\phi_i})$ for every $\ket{\phi_i}$ \cite{Wootters-1998a}.  Hence, if $2(1+\kappa)<\tr[\rho\mc{B}]=\sum_{i}p_i\bra{\phi_i}\rho\ket{\phi_i}$, then there must exist at least one pure state that violates the CHSH Inequality having the same concurrence as $\rho$.  We thus summarize our findings in the following theorem, with the first part recovering a previous result of  Refs. \cite{Cabello-2007a, Brunner-2007a}.
\begin{theorem}
\label{thm1}
In a two-qubit CHSH experiment with detection efficiency $\eta$ for one of the parties and perfect efficiency for the other, there exists measurement directions $\{\hat{a}_0,\hat{a}_1,\hat{b}_0,\hat{b}_1\}$ and an entangled state $\rho$ such that $\tr[\rho\mc{B}]>2$ if $\eta>1/2$.  Moreover, if $\tr[\rho\mc{B}]>2(1+\kappa)$ for any $\kappa\geq 0$, then
\begin{align}
\label{Eq:main_bound}
C(\rho)^2 &< \frac{(2\eta-1)(\kappa(2+\kappa)+2\eta(1-\eta))}{(1-2\eta+2\eta^2)^2}\notag\\
&\;\;+\frac{2\eta(1-\eta)(1+\kappa)\sqrt{1-\kappa(2+\kappa)-4\eta(1-\eta)}}{(1-2\eta+2\eta^2)^2},
\end{align}
where $C(\rho)$ is the concurrence of $\rho$.
\end{theorem}
\begin{figure}[t]
\centering
\includegraphics[scale=.47]{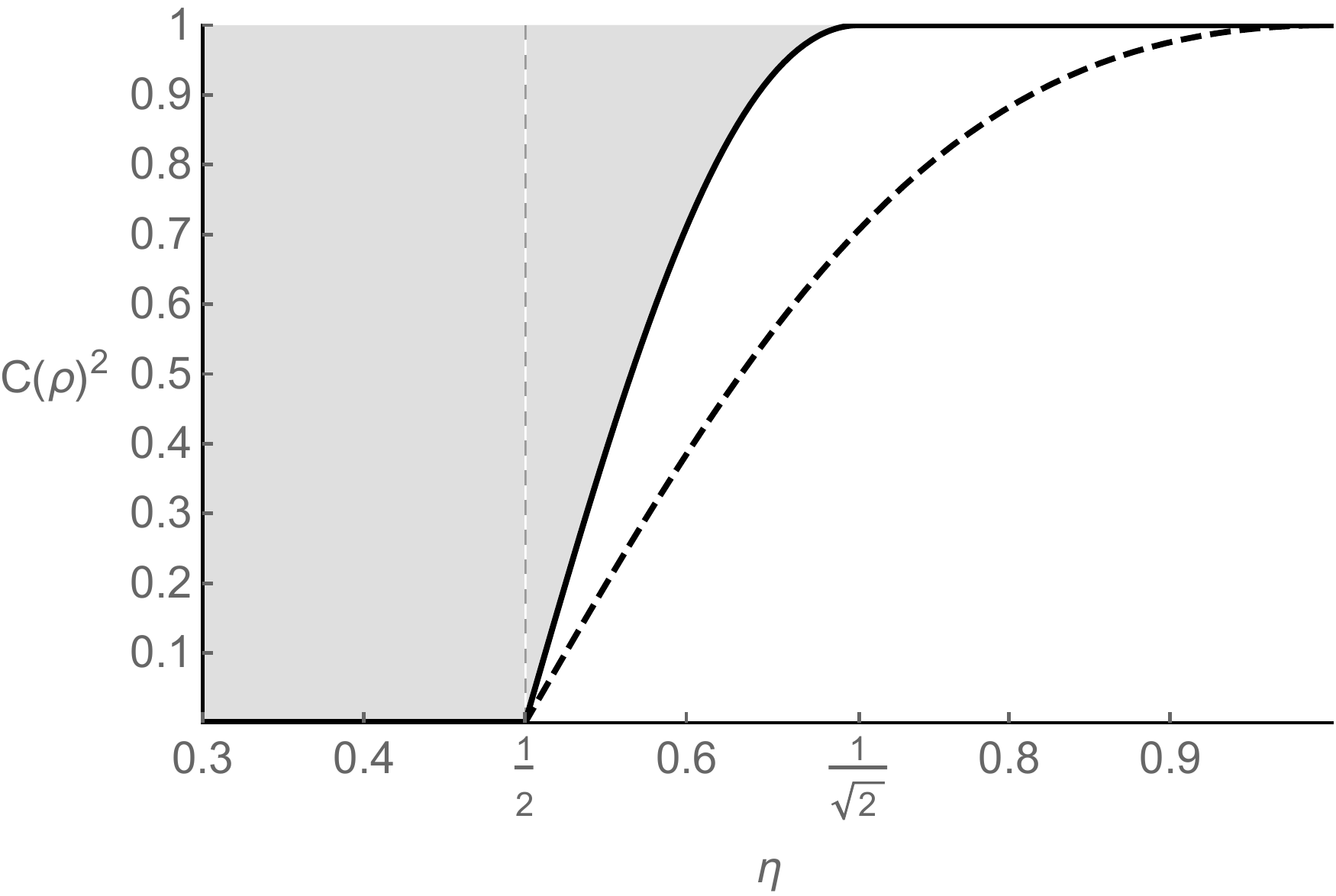}
\caption{For a given detection efficiency $\eta$, a state with concurrence-squared at or above the solid line will never violate the CHSH inequality.  On the other hand, for $\eta>1/2$, there always exists a state that can violate the CHSH Inequality whose concurrence-squared is given by the dashed line.} 
\label{Fig:Upper_and_Lower_Bounds}
\end{figure}
\noindent From Eq. \eqref{Eq:main_bound}, the Bell operator $\mc{B}$ constructed above can be seen as both a witness of nonlocality and a witness for the amount of entanglement in a state.  Note that Eq. \eqref{Eq:main_bound} provides a tight bound when $\eta=1$; in this case the bound becomes $C(\rho)^2<\kappa(2+\kappa)$, and for any pure state satisfying $C(\rho)^2=\kappa(2+\kappa)$ suitable measurement directions exist for which $\tr[\rho\beta]=2(1+\kappa)$ \cite{Horodecki-1995a, Verstraete-2002a}.  We plot the $\kappa=0$ bound (i.e. Eq. \eqref{Eq:violation_ub}) in Fig. \ref{Fig:Upper_and_Lower_Bounds}.  Observe the similarity to Fig. 1 in Ref. \cite{Liang-2011a} for the entanglement in states violating the CHSH Inequality with two-sided detection inefficiency. 

An interesting question is whether the converse to the first part of Theorem \ref{thm1} is also true.  That is, for $\eta\leq 1/2$, can the CHSH Inequality still be violated?  By recalling the approach of Massar and Pironio \cite{Massar-2003a}, we can show that this is not possible by explicitly constructing a LHV model that simulates any family of no-signaling distributions $p(a,b|x,y)$  provided Bob is allowed to report an additional outcome $\emptyset$ with probability $1-\eta$ for each of his inputs.  The no-signaling restriction on $p(a,b|x,y)$ says that $\sum_{a}p(a,b|x,y)=\sum_a p(a,b|x',y)$ for all $b,x,x',y$, and $\sum_bp(a,b|x,y)=\sum_bp(a,b|x,y')$ for $a,x,y,y'$.  Here the variables belong to discrete sets $a\in\mc{A}$, $b\in\mc{B}$, $x\in\mc{X}$, and $y\in\mc{Y}$.  

The local model is defined by a shared variable $\lambda$, with distribution $q(\lambda)$, along with local conditional distributions $q(a|x,\lambda)$ and $q(b|y,\lambda)$ where $a\in\mc{A}$, $b\in\mc{B}\cup\{\emptyset\}$, $x\in\mc{X}$, and $y\in\mc{Y}$.  The joint distribution is given by 
\begin{equation}
\label{Eq:LHV_Model}
q(a,b|x,y)=\sum_{\lambda} q(\lambda) q(a|x,\lambda) q(b|y,\lambda),
\end{equation}
and a successful simulation satisfies $q(a,b|x,y)=\eta p(a,b|x,y)$ for $b\not=\emptyset$ and $q(a,\emptyset|x,y)=(1-\eta)p(a|x)$. The idea of Massar and Pironio involves the shared variable $\lambda=\{(b',y')\}_{b'\in\mc{B},y'\in\mc{Y}}$ with distribution $q(\lambda)= p(b'|y')/|\mc{Y}|$.  Bob's local channel is given by the distributions $q(b|y,\lambda)=\delta_{bb'}\delta_{yy'}$ for $b\not=\emptyset$ and $q(b|y,\lambda)=1-\delta_{yy'}$ for $b=\emptyset$, while Alice's channel is given by $q(a|x,\lambda) = p(ab'|xy')/p(b'|y')$.  Substituting these into Eq. \eqref{Eq:LHV_Model} yields a correct simulation provided that $\eta=1/|\mc{Y}|$. For any $\eta<1/|\mc{Y}|$, the above model is modified by having Bob relabel each output $b\to\emptyset$ with a certain probability that is the same for each outcome $b\not=\emptyset$.  Thus, when $|\mc{Y}|=2$ the local model simulates the correct correlations for all $\eta\leq 1/2$, as shown in Fig. \ref{Fig:Upper_and_Lower_Bounds}.


The second scenario we consider involves detection inefficiency for only one of Bob's measurement choices.  In this case, only one of Bob's POVMs has the form of Eqns. \eqref{Eq:Bob-POVM1}-\eqref{Eq:Bob-POVM2}, say its $y=1$, while the choice $y=0$ remains a projective measurement of Eq. \eqref{Eq:Proj-Bob}.  The new Bell operator is given by
\begin{align}
\mc{B}'=\sum_{i,j\in\{x,z\}}c_{ij}\sigma_i\otimes\sigma_j+(r_x\sigma_x+r_z\sigma_z)\otimes\mbb{I}
\end{align}
where
\begin{align}
c_{xx}&=-\eta\cos\theta_A\cos\theta_B&c_{xz}&=\cos\theta_A(1-\eta\sin\theta_B)\notag\\
c_{zx}&=\eta\cos\theta_B(1-\sin\theta_A)&c_{zz}&=1+\eta\sin\theta_B\notag\\
&&&\;\;+\sin\theta_A(1-\eta\sin\theta_B)\notag\\
r_x&=-(1-\eta)\cos\theta_A&r_z&=(1-\eta)(1-\sin\theta_A).\notag
\end{align}
Again due to its simplified structure, the eigenvalues of $\mc{B}'$ can be explicitly computed.  The largest eigenvalue $\lambda_{max}$ of $\mc{B}'$ is found to satisfy $\lambda^2_{max}=r_x^2+r_z^2+c_{xx}^2+c_{xz}^2+c_{zx}^2+c_{zz}^2+2\{(r_xc_{xx}+r_zc_{zx})^2+(r_xc_{xz}+r_zc_{zz})^2+(c_{xx}c_{zz}-c_{xz}c_{zx})^2\}^{1/2}$.  Substituting in the matrix components of $\mc{B}'$ and performing some algebraic simplifications, we have
\begin{align}
\lambda^2_{max}&=4(1-\eta(1-\eta)(1-\sin\theta_A)+4\{\eta^2(1-\sin\theta_A)\notag\\
&\;\times[\cos^2\theta_B(1+\sin\theta_A)+(1-\eta)^2(1-\sin\theta_A)]\}^{1/2}\notag\\
&\geq 4,
\end{align}
with equality holding if and only if $\eta(1-\sin\theta_A)=0$ or $\cos^2\theta_B(1+\sin\theta_A)=0$.  We thus arrive at the second main result of this paper.
\begin{proposition}
\label{prop1}
In a two-qubit CHSH experiment with detection efficiency $\eta$ for one choice of measurement for one of the parties and perfect efficiency for all others, there exists a state $\rho$ such that $\tr[\rho\mc{B}']>2$ for any $\eta>0$ and any set of measurement directions $\{\hat{a}_0,\hat{a}_1,\hat{b}_0,\hat{b}_1\}$ with $\hat{a}_0\not=\pm\hat{a}_1$ and $\hat{b}_0\not=\pm\hat{b}_1$.
\end{proposition}

It is informative to compare this proposition with Thm. \ref{thm1}.  The latter just guarantees the existence of some measurement directions that generate nonlocal correlations for $\eta>1/2$.  In contrast, Prop. \ref{prop1} says that \textit{any} measurement directions will do.  Concerning entanglement, Thm. \ref{thm1} places an upper bound on the entanglement needed to violate the CHSH Inequality.  One may wonder if likewise the entanglement in states violating the CHSH Inequality for $\mc{B}'$ goes to zero as $\eta\to 0$.  Interestingly, this turns out not to be the case.  The reason is that as $\eta\to 0$, $\mc{B}'$ develops a double degeneracy in its largest eigenvalue $\lambda_{\max}=2$.  Hence, the associated eigenspace is two-dimensional, and thus one can take an entangled superposition of eigenstates and still violate the CHSH Inequality for $\eta$ very close to zero.  The greatest amount of entanglement that can be generated by taking such a superposition depends on the particular choices of $\theta_A$ and $\theta_B$.  For example, when $\theta_A=\theta_B=0$ and $\eta \rightarrow 0$, a state with entanglement $C(\op{\phi}{\phi})=1/\sqrt{2}$ can violate the CHSH Inequality for arbitrary $\eta>0$.  Nevertheless, by performing an analysis similar to the one given above, one can show that $\kappa_{\max}$, the largest squared Schmidt coefficient of $\ket{\phi}$, is upper bounded by
\begin{align}
&\dfrac{\sqrt{2} - \sqrt{1 + \eta^2 +2 \eta \cos\theta_A \cos\theta_B + (1-\eta^2) \sin\theta_A}}{2(1 - \eta) \sqrt{1 - \sin\theta_A}}+\dfrac{1}{2}\notag
\end{align}
whenever $\bra{\phi}\mc{B}'\ket{\phi}>2$.  In the limit $\eta\to 0$ and $\theta_A\to-\pi/2^+$, the RHS converges to one.  Hence as Alice's measurement directions become closer to being parallel, a smaller amount of entanglement is necessary to demonstrate nonlocality with very poor detection efficiency.

We finally note that Prop. \ref{prop1} also has implications for the relationship between nonlocality and measurement incompatibility \cite{Fine-1982a, Bene-2018a, Hirsch-2018a}.  Wolf \textit{et. al} has shown that for any two non-commuting observables on Alice's side, nonlocality can always be demonstrated using arbitrary non-commuting observables on Bob's side \cite{Wolf-2009a}.  Our results provide a slight generalization of this in two-qubit systems.  According to Prop. \ref{prop1}, for any two non-commuting observables on Alice's side, nonlocality can always be demonstrated using incompatible POVMs on Bob's side in which one of them is any standard observable and the other is any non-commuting ``coarse-grained'' observable, i.e. having the form of Eqns. \eqref{Eq:Bob-POVM1}-\eqref{Eq:Bob-POVM2}. 


In this paper we have considered two CHSH experiments in which Bob has inefficient detectors for either both or just one of his measurement choices.  Our work can be seen as studying extreme cases in a more general scenario where Alice and Bob have detection efficiencies $\eta^A_x$ and $\eta^B_y$ for each measurement choice $x$ and $y$, respectively.  When $\eta^A_0=\eta^A_1=\eta^B_0=\eta^B_1=\eta$ the locality bound is $\eta=2/3$, i.e. the threshold in which any quantum correlations can be simulated by a LHV model \cite{Eberhard-1993a, Massar-2003a}.  When $\eta^A_0=\eta^A_1=1$ and $\eta^B_1=\eta^B_2=\eta$, the locality bound is $\eta=1/2$, and when $\eta^A_0=\eta^A_1=\eta^B_0=1$, Prop. \ref{prop1} shows there is no locality bound for any $\eta^B_2>0$.  It would be interesting to study the locality bound for general values of $(\eta^A_0,\eta^A_1,\eta^B_0,\eta^B_1)$.

\medskip

\noindent\textbf{\textit{Acknowledgments.}}  
We are very grateful to Nicolas Brunner, Yeong-Cherng Liang, and Daniel Cavalcanti for pointing out previous work on asymmetric Bell tests.  We are also thankful to Jed Kaniewski for constructive comments on the content of Theorem 1.   This work is supported by the National Science Foundation (NSF) Early CAREER Award No. 1352326.

\bibliography{NonlocalityBib}

\begin{thebibliography}{33}%
\makeatletter
\providecommand \@ifxundefined [1]{%
 \@ifx{#1\undefined}
}%
\providecommand \@ifnum [1]{%
 \ifnum #1\expandafter \@firstoftwo
 \else \expandafter \@secondoftwo
 \fi
}%
\providecommand \@ifx [1]{%
 \ifx #1\expandafter \@firstoftwo
 \else \expandafter \@secondoftwo
 \fi
}%
\providecommand \natexlab [1]{#1}%
\providecommand \enquote  [1]{``#1''}%
\providecommand \bibnamefont  [1]{#1}%
\providecommand \bibfnamefont [1]{#1}%
\providecommand \citenamefont [1]{#1}%
\providecommand \href@noop [0]{\@secondoftwo}%
\providecommand \href [0]{\begingroup \@sanitize@url \@href}%
\providecommand \@href[1]{\@@startlink{#1}\@@href}%
\providecommand \@@href[1]{\endgroup#1\@@endlink}%
\providecommand \@sanitize@url [0]{\catcode `\\12\catcode `\$12\catcode
  `\&12\catcode `\#12\catcode `\^12\catcode `\_12\catcode `\%12\relax}%
\providecommand \@@startlink[1]{}%
\providecommand \@@endlink[0]{}%
\providecommand \url  [0]{\begingroup\@sanitize@url \@url }%
\providecommand \@url [1]{\endgroup\@href {#1}{\urlprefix }}%
\providecommand \urlprefix  [0]{URL }%
\providecommand \Eprint [0]{\href }%
\providecommand \doibase [0]{http://dx.doi.org/}%
\providecommand \selectlanguage [0]{\@gobble}%
\providecommand \bibinfo  [0]{\@secondoftwo}%
\providecommand \bibfield  [0]{\@secondoftwo}%
\providecommand \translation [1]{[#1]}%
\providecommand \BibitemOpen [0]{}%
\providecommand \bibitemStop [0]{}%
\providecommand \bibitemNoStop [0]{.\EOS\space}%
\providecommand \EOS [0]{\spacefactor3000\relax}%
\providecommand \BibitemShut  [1]{\csname bibitem#1\endcsname}%
\let\auto@bib@innerbib\@empty
\bibitem [{\citenamefont {Werner}(1989)}]{Werner-1989a}%
  \BibitemOpen
  \bibfield  {author} {\bibinfo {author} {\bibfnamefont {R.~F.}\ \bibnamefont
  {Werner}},\ }\href {\doibase 10.1103/PhysRevA.40.4277} {\bibfield  {journal}
  {\bibinfo  {journal} {Phys. Rev. A}\ }\textbf {\bibinfo {volume} {40}},\
  \bibinfo {pages} {4277} (\bibinfo {year} {1989})}\BibitemShut {NoStop}%
\bibitem [{\citenamefont {Barrett}(2002)}]{Barrett-2002a}%
  \BibitemOpen
  \bibfield  {author} {\bibinfo {author} {\bibfnamefont {J.}~\bibnamefont
  {Barrett}},\ }\href {\doibase 10.1103/PhysRevA.65.042302} {\bibfield
  {journal} {\bibinfo  {journal} {Phys. Rev. A}\ }\textbf {\bibinfo {volume}
  {65}},\ \bibinfo {pages} {042302} (\bibinfo {year} {2002})}\BibitemShut
  {NoStop}%
\bibitem [{\citenamefont {Bowles}\ \emph {et~al.}(2016)\citenamefont {Bowles},
  \citenamefont {Francfort}, \citenamefont {Fillettaz}, \citenamefont
  {Hirsch},\ and\ \citenamefont {Brunner}}]{Bowles-2016a}%
  \BibitemOpen
  \bibfield  {author} {\bibinfo {author} {\bibfnamefont {J.}~\bibnamefont
  {Bowles}}, \bibinfo {author} {\bibfnamefont {J.}~\bibnamefont {Francfort}},
  \bibinfo {author} {\bibfnamefont {M.}~\bibnamefont {Fillettaz}}, \bibinfo
  {author} {\bibfnamefont {F.}~\bibnamefont {Hirsch}}, \ and\ \bibinfo {author}
  {\bibfnamefont {N.}~\bibnamefont {Brunner}},\ }\href {\doibase
  10.1103/PhysRevLett.116.130401} {\bibfield  {journal} {\bibinfo  {journal}
  {Phys. Rev. Lett.}\ }\textbf {\bibinfo {volume} {116}},\ \bibinfo {pages}
  {130401} (\bibinfo {year} {2016})}\BibitemShut {NoStop}%
\bibitem [{\citenamefont {Bell}(1964)}]{Bell-1964a}%
  \BibitemOpen
  \bibfield  {author} {\bibinfo {author} {\bibfnamefont {J.~S.}\ \bibnamefont
  {Bell}},\ }\href@noop {} {\bibfield  {journal} {\bibinfo  {journal}
  {Physics}\ }\textbf {\bibinfo {volume} {1}},\ \bibinfo {pages} {195}
  (\bibinfo {year} {1964})}\BibitemShut {NoStop}%
\bibitem [{\citenamefont {Horodecki}\ \emph {et~al.}(2009)\citenamefont
  {Horodecki}, \citenamefont {Horodecki}, \citenamefont {Horodecki},\ and\
  \citenamefont {Horodecki}}]{Horodecki-2009a}%
  \BibitemOpen
  \bibfield  {author} {\bibinfo {author} {\bibfnamefont {R.}~\bibnamefont
  {Horodecki}}, \bibinfo {author} {\bibfnamefont {P.}~\bibnamefont
  {Horodecki}}, \bibinfo {author} {\bibfnamefont {M.}~\bibnamefont
  {Horodecki}}, \ and\ \bibinfo {author} {\bibfnamefont {K.}~\bibnamefont
  {Horodecki}},\ }\href {\doibase 10.1103/RevModPhys.81.865} {\bibfield
  {journal} {\bibinfo  {journal} {Rev. Mod. Phys.}\ }\textbf {\bibinfo {volume}
  {81}},\ \bibinfo {eid} {865} (\bibinfo {year} {2009})}\BibitemShut {NoStop}%
\bibitem [{\citenamefont {Eberhard}(1993)}]{Eberhard-1993a}%
  \BibitemOpen
  \bibfield  {author} {\bibinfo {author} {\bibfnamefont {P.~H.}\ \bibnamefont
  {Eberhard}},\ }\href {\doibase 10.1103/PhysRevA.47.R747} {\bibfield
  {journal} {\bibinfo  {journal} {Phys. Rev. A}\ }\textbf {\bibinfo {volume}
  {47}},\ \bibinfo {pages} {R747} (\bibinfo {year} {1993})}\BibitemShut
  {NoStop}%
\bibitem [{\citenamefont {Hardy}(1993)}]{Hardy-1993a}%
  \BibitemOpen
  \bibfield  {author} {\bibinfo {author} {\bibfnamefont {L.}~\bibnamefont
  {Hardy}},\ }\href {\doibase 10.1103/PhysRevLett.71.1665} {\bibfield
  {journal} {\bibinfo  {journal} {Phys. Rev. Lett.}\ }\textbf {\bibinfo
  {volume} {71}},\ \bibinfo {pages} {1665} (\bibinfo {year}
  {1993})}\BibitemShut {NoStop}%
\bibitem [{\citenamefont {Ac\'{\i}n}\ \emph {et~al.}(2002)\citenamefont
  {Ac\'{\i}n}, \citenamefont {Durt}, \citenamefont {Gisin},\ and\ \citenamefont
  {Latorre}}]{Acin-2002a}%
  \BibitemOpen
  \bibfield  {author} {\bibinfo {author} {\bibfnamefont {A.}~\bibnamefont
  {Ac\'{\i}n}}, \bibinfo {author} {\bibfnamefont {T.}~\bibnamefont {Durt}},
  \bibinfo {author} {\bibfnamefont {N.}~\bibnamefont {Gisin}}, \ and\ \bibinfo
  {author} {\bibfnamefont {J.~I.}\ \bibnamefont {Latorre}},\ }\href {\doibase
  10.1103/PhysRevA.65.052325} {\bibfield  {journal} {\bibinfo  {journal} {Phys.
  Rev. A}\ }\textbf {\bibinfo {volume} {65}},\ \bibinfo {pages} {052325}
  (\bibinfo {year} {2002})}\BibitemShut {NoStop}%
\bibitem [{\citenamefont {Brunner}\ \emph {et~al.}(2005)\citenamefont
  {Brunner}, \citenamefont {Gisin},\ and\ \citenamefont
  {Scarani}}]{Brunner-2005a}%
  \BibitemOpen
  \bibfield  {author} {\bibinfo {author} {\bibfnamefont {N.}~\bibnamefont
  {Brunner}}, \bibinfo {author} {\bibfnamefont {N.}~\bibnamefont {Gisin}}, \
  and\ \bibinfo {author} {\bibfnamefont {V.}~\bibnamefont {Scarani}},\ }\href
  {\doibase 10.1088/1367-2630/7/1/088} {\bibfield  {journal} {\bibinfo
  {journal} {New Journal of Physics}\ }\textbf {\bibinfo {volume} {7}},\
  \bibinfo {pages} {88} (\bibinfo {year} {2005})}\BibitemShut {NoStop}%
\bibitem [{\citenamefont {Cabello}\ and\ \citenamefont
  {Larsson}(2007)}]{Cabello-2007a}%
  \BibitemOpen
  \bibfield  {author} {\bibinfo {author} {\bibfnamefont {A.}~\bibnamefont
  {Cabello}}\ and\ \bibinfo {author} {\bibfnamefont {J.-A.}\ \bibnamefont
  {Larsson}},\ }\href {\doibase 10.1103/PhysRevLett.98.220402} {\bibfield
  {journal} {\bibinfo  {journal} {Phys. Rev. Lett.}\ }\textbf {\bibinfo
  {volume} {98}},\ \bibinfo {pages} {220402} (\bibinfo {year}
  {2007})}\BibitemShut {NoStop}%
\bibitem [{\citenamefont {Brunner}\ \emph {et~al.}(2007)\citenamefont
  {Brunner}, \citenamefont {Gisin}, \citenamefont {Scarani},\ and\
  \citenamefont {Simon}}]{Brunner-2007a}%
  \BibitemOpen
  \bibfield  {author} {\bibinfo {author} {\bibfnamefont {N.}~\bibnamefont
  {Brunner}}, \bibinfo {author} {\bibfnamefont {N.}~\bibnamefont {Gisin}},
  \bibinfo {author} {\bibfnamefont {V.}~\bibnamefont {Scarani}}, \ and\
  \bibinfo {author} {\bibfnamefont {C.}~\bibnamefont {Simon}},\ }\href
  {\doibase 10.1103/PhysRevLett.98.220403} {\bibfield  {journal} {\bibinfo
  {journal} {Phys. Rev. Lett.}\ }\textbf {\bibinfo {volume} {98}},\ \bibinfo
  {pages} {220403} (\bibinfo {year} {2007})}\BibitemShut {NoStop}%
\bibitem [{\citenamefont {P\'al}\ and\ \citenamefont
  {V\'ertesi}(2010)}]{Pal-2010a}%
  \BibitemOpen
  \bibfield  {author} {\bibinfo {author} {\bibfnamefont {K.~F.}\ \bibnamefont
  {P\'al}}\ and\ \bibinfo {author} {\bibfnamefont {T.}~\bibnamefont
  {V\'ertesi}},\ }\href {\doibase 10.1103/PhysRevA.82.022116} {\bibfield
  {journal} {\bibinfo  {journal} {Phys. Rev. A}\ }\textbf {\bibinfo {volume}
  {82}},\ \bibinfo {pages} {022116} (\bibinfo {year} {2010})}\BibitemShut
  {NoStop}%
\bibitem [{\citenamefont {Vidick}\ and\ \citenamefont
  {Wehner}(2011)}]{Vidick-2011a}%
  \BibitemOpen
  \bibfield  {author} {\bibinfo {author} {\bibfnamefont {T.}~\bibnamefont
  {Vidick}}\ and\ \bibinfo {author} {\bibfnamefont {S.}~\bibnamefont
  {Wehner}},\ }\href {\doibase 10.1103/PhysRevA.83.052310} {\bibfield
  {journal} {\bibinfo  {journal} {Phys. Rev. A}\ }\textbf {\bibinfo {volume}
  {83}},\ \bibinfo {pages} {052310} (\bibinfo {year} {2011})}\BibitemShut
  {NoStop}%
\bibitem [{\citenamefont {Junge}\ and\ \citenamefont
  {Palazuelos}(2011)}]{Junge-2011a}%
  \BibitemOpen
  \bibfield  {author} {\bibinfo {author} {\bibfnamefont {M.}~\bibnamefont
  {Junge}}\ and\ \bibinfo {author} {\bibfnamefont {C.}~\bibnamefont
  {Palazuelos}},\ }\href {\doibase 10.1007/s00220-011-1296-8} {\bibfield
  {journal} {\bibinfo  {journal} {Communications in Mathematical Physics}\
  }\textbf {\bibinfo {volume} {306}},\ \bibinfo {pages} {695} (\bibinfo {year}
  {2011})}\BibitemShut {NoStop}%
\bibitem [{\citenamefont {Liang}\ \emph {et~al.}(2011)\citenamefont {Liang},
  \citenamefont {V\'ertesi},\ and\ \citenamefont {Brunner}}]{Liang-2011a}%
  \BibitemOpen
  \bibfield  {author} {\bibinfo {author} {\bibfnamefont {Y.-C.}\ \bibnamefont
  {Liang}}, \bibinfo {author} {\bibfnamefont {T.}~\bibnamefont {V\'ertesi}}, \
  and\ \bibinfo {author} {\bibfnamefont {N.}~\bibnamefont {Brunner}},\ }\href
  {\doibase 10.1103/PhysRevA.83.022108} {\bibfield  {journal} {\bibinfo
  {journal} {Phys. Rev. A}\ }\textbf {\bibinfo {volume} {83}},\ \bibinfo
  {pages} {022108} (\bibinfo {year} {2011})}\BibitemShut {NoStop}%
\bibitem [{\citenamefont {Vallone}\ \emph {et~al.}(2014)\citenamefont
  {Vallone}, \citenamefont {Lima}, \citenamefont {G\'omez}, \citenamefont
  {Ca\~nas}, \citenamefont {Larsson}, \citenamefont {Mataloni},\ and\
  \citenamefont {Cabello}}]{Vallone-2014a}%
  \BibitemOpen
  \bibfield  {author} {\bibinfo {author} {\bibfnamefont {G.}~\bibnamefont
  {Vallone}}, \bibinfo {author} {\bibfnamefont {G.}~\bibnamefont {Lima}},
  \bibinfo {author} {\bibfnamefont {E.~S.}\ \bibnamefont {G\'omez}}, \bibinfo
  {author} {\bibfnamefont {G.}~\bibnamefont {Ca\~nas}}, \bibinfo {author}
  {\bibfnamefont {J.-A.}\ \bibnamefont {Larsson}}, \bibinfo {author}
  {\bibfnamefont {P.}~\bibnamefont {Mataloni}}, \ and\ \bibinfo {author}
  {\bibfnamefont {A.}~\bibnamefont {Cabello}},\ }\href {\doibase
  10.1103/PhysRevA.89.012102} {\bibfield  {journal} {\bibinfo  {journal} {Phys.
  Rev. A}\ }\textbf {\bibinfo {volume} {89}},\ \bibinfo {pages} {012102}
  (\bibinfo {year} {2014})}\BibitemShut {NoStop}%
\bibitem [{\citenamefont {Palazuelos}(2014)}]{Palazuelos-2014a}%
  \BibitemOpen
  \bibfield  {author} {\bibinfo {author} {\bibfnamefont {C.}~\bibnamefont
  {Palazuelos}},\ }\href {\doibase https://doi.org/10.1016/j.jfa.2014.07.028}
  {\bibfield  {journal} {\bibinfo  {journal} {Journal of Functional Analysis}\
  }\textbf {\bibinfo {volume} {267}},\ \bibinfo {pages} {1959 } (\bibinfo
  {year} {2014})}\BibitemShut {NoStop}%
\bibitem [{\citenamefont {Christensen}\ \emph {et~al.}(2015)\citenamefont
  {Christensen}, \citenamefont {Liang}, \citenamefont {Brunner}, \citenamefont
  {Gisin},\ and\ \citenamefont {Kwiat}}]{Christensen-2015a}%
  \BibitemOpen
  \bibfield  {author} {\bibinfo {author} {\bibfnamefont {B.~G.}\ \bibnamefont
  {Christensen}}, \bibinfo {author} {\bibfnamefont {Y.-C.}\ \bibnamefont
  {Liang}}, \bibinfo {author} {\bibfnamefont {N.}~\bibnamefont {Brunner}},
  \bibinfo {author} {\bibfnamefont {N.}~\bibnamefont {Gisin}}, \ and\ \bibinfo
  {author} {\bibfnamefont {P.~G.}\ \bibnamefont {Kwiat}},\ }\href {\doibase
  10.1103/PhysRevX.5.041052} {\bibfield  {journal} {\bibinfo  {journal} {Phys.
  Rev. X}\ }\textbf {\bibinfo {volume} {5}},\ \bibinfo {pages} {041052}
  (\bibinfo {year} {2015})}\BibitemShut {NoStop}%
\bibitem [{\citenamefont {A.~A.~M\'{e}thot}(2007)}]{Methot-2007a}%
  \BibitemOpen
  \bibfield  {author} {\bibinfo {author} {\bibfnamefont {V.~S.}\ \bibnamefont
  {A.~A.~M\'{e}thot}},\ }\href@noop {} {\bibfield  {journal} {\bibinfo
  {journal} {Quant. Inf. Comput.}\ }\textbf {\bibinfo {volume} {7}},\ \bibinfo
  {pages} {157} (\bibinfo {year} {2007})},\ \Eprint
  {http://arxiv.org/abs/quant-ph/0601210} {quant-ph/0601210} \BibitemShut
  {NoStop}%
\bibitem [{\citenamefont {Clauser}\ \emph {et~al.}(1969)\citenamefont
  {Clauser}, \citenamefont {Horne}, \citenamefont {Shimony},\ and\
  \citenamefont {Holt}}]{Clauser-1969a}%
  \BibitemOpen
  \bibfield  {author} {\bibinfo {author} {\bibfnamefont {J.~F.}\ \bibnamefont
  {Clauser}}, \bibinfo {author} {\bibfnamefont {M.~A.}\ \bibnamefont {Horne}},
  \bibinfo {author} {\bibfnamefont {A.}~\bibnamefont {Shimony}}, \ and\
  \bibinfo {author} {\bibfnamefont {R.~A.}\ \bibnamefont {Holt}},\ }\href
  {\doibase 10.1103/PhysRevLett.23.880} {\bibfield  {journal} {\bibinfo
  {journal} {Phys. Rev. Lett.}\ }\textbf {\bibinfo {volume} {23}},\ \bibinfo
  {pages} {880} (\bibinfo {year} {1969})}\BibitemShut {NoStop}%
\bibitem [{\citenamefont {Pearle}(1970)}]{Pearle-1970a}%
  \BibitemOpen
  \bibfield  {author} {\bibinfo {author} {\bibfnamefont {P.~M.}\ \bibnamefont
  {Pearle}},\ }\href {\doibase 10.1103/PhysRevD.2.1418} {\bibfield  {journal}
  {\bibinfo  {journal} {Phys. Rev. D}\ }\textbf {\bibinfo {volume} {2}},\
  \bibinfo {pages} {1418} (\bibinfo {year} {1970})}\BibitemShut {NoStop}%
\bibitem [{\citenamefont {Brunner}\ \emph {et~al.}(2014)\citenamefont
  {Brunner}, \citenamefont {Cavalcanti}, \citenamefont {Pironio}, \citenamefont
  {Scarani},\ and\ \citenamefont {Wehner}}]{Brunner-2014a}%
  \BibitemOpen
  \bibfield  {author} {\bibinfo {author} {\bibfnamefont {N.}~\bibnamefont
  {Brunner}}, \bibinfo {author} {\bibfnamefont {D.}~\bibnamefont {Cavalcanti}},
  \bibinfo {author} {\bibfnamefont {S.}~\bibnamefont {Pironio}}, \bibinfo
  {author} {\bibfnamefont {V.}~\bibnamefont {Scarani}}, \ and\ \bibinfo
  {author} {\bibfnamefont {S.}~\bibnamefont {Wehner}},\ }\href {\doibase
  10.1103/RevModPhys.86.419} {\bibfield  {journal} {\bibinfo  {journal} {Rev.
  Mod. Phys.}\ }\textbf {\bibinfo {volume} {86}},\ \bibinfo {pages} {419}
  (\bibinfo {year} {2014})}\BibitemShut {NoStop}%
\bibitem [{\citenamefont {Clauser}\ and\ \citenamefont
  {Horne}(1974)}]{Clauser-1974a}%
  \BibitemOpen
  \bibfield  {author} {\bibinfo {author} {\bibfnamefont {J.~F.}\ \bibnamefont
  {Clauser}}\ and\ \bibinfo {author} {\bibfnamefont {M.~A.}\ \bibnamefont
  {Horne}},\ }\href {\doibase 10.1103/PhysRevD.10.526} {\bibfield  {journal}
  {\bibinfo  {journal} {Phys. Rev. D}\ }\textbf {\bibinfo {volume} {10}},\
  \bibinfo {pages} {526} (\bibinfo {year} {1974})}\BibitemShut {NoStop}%
\bibitem [{\citenamefont {Cirel'son}(1980)}]{Cirel'son-1980a}%
  \BibitemOpen
  \bibfield  {author} {\bibinfo {author} {\bibfnamefont {B.~S.}\ \bibnamefont
  {Cirel'son}},\ }\href {\doibase 10.1007/BF00417500} {\bibfield  {journal}
  {\bibinfo  {journal} {Letters in Mathematical Physics}\ }\textbf {\bibinfo
  {volume} {4}},\ \bibinfo {pages} {93} (\bibinfo {year} {1980})}\BibitemShut
  {NoStop}%
\bibitem [{\citenamefont {Wootters}(1998)}]{Wootters-1998a}%
  \BibitemOpen
  \bibfield  {author} {\bibinfo {author} {\bibfnamefont {W.~K.}\ \bibnamefont
  {Wootters}},\ }\href {\doibase 10.1103/PhysRevLett.80.2245} {\bibfield
  {journal} {\bibinfo  {journal} {Phys. Rev. Lett.}\ }\textbf {\bibinfo
  {volume} {80}},\ \bibinfo {pages} {2245} (\bibinfo {year}
  {1998})}\BibitemShut {NoStop}%
\bibitem [{\citenamefont {Verstraete}\ and\ \citenamefont
  {Wolf}(2002)}]{Verstraete-2002a}%
  \BibitemOpen
  \bibfield  {author} {\bibinfo {author} {\bibfnamefont {F.}~\bibnamefont
  {Verstraete}}\ and\ \bibinfo {author} {\bibfnamefont {M.~M.}\ \bibnamefont
  {Wolf}},\ }\href {\doibase 10.1103/PhysRevLett.89.170401} {\bibfield
  {journal} {\bibinfo  {journal} {Phys. Rev. Lett.}\ }\textbf {\bibinfo
  {volume} {89}},\ \bibinfo {pages} {170401} (\bibinfo {year}
  {2002})}\BibitemShut {NoStop}%
\bibitem [{\citenamefont {Lasserre}(1995)}]{Lasserre-1995a}%
  \BibitemOpen
  \bibfield  {author} {\bibinfo {author} {\bibfnamefont {J.~B.}\ \bibnamefont
  {Lasserre}},\ }\href {\doibase 10.1109/9.402252} {\bibfield  {journal}
  {\bibinfo  {journal} {IEEE Transactions on Automatic Control}\ }\textbf
  {\bibinfo {volume} {40}},\ \bibinfo {pages} {1500} (\bibinfo {year}
  {1995})}\BibitemShut {NoStop}%
\bibitem [{\citenamefont {Horodecki}\ \emph {et~al.}(1995)\citenamefont
  {Horodecki}, \citenamefont {Horodecki},\ and\ \citenamefont
  {Horodecki}}]{Horodecki-1995a}%
  \BibitemOpen
  \bibfield  {author} {\bibinfo {author} {\bibfnamefont {R.}~\bibnamefont
  {Horodecki}}, \bibinfo {author} {\bibfnamefont {P.}~\bibnamefont
  {Horodecki}}, \ and\ \bibinfo {author} {\bibfnamefont {M.}~\bibnamefont
  {Horodecki}},\ }\href {\doibase
  http://dx.doi.org/10.1016/0375-9601(95)00214-N} {\bibfield  {journal}
  {\bibinfo  {journal} {Physics Letters A}\ }\textbf {\bibinfo {volume}
  {200}},\ \bibinfo {pages} {340 } (\bibinfo {year} {1995})}\BibitemShut
  {NoStop}%
\bibitem [{\citenamefont {Massar}\ and\ \citenamefont
  {Pironio}(2003)}]{Massar-2003a}%
  \BibitemOpen
  \bibfield  {author} {\bibinfo {author} {\bibfnamefont {S.}~\bibnamefont
  {Massar}}\ and\ \bibinfo {author} {\bibfnamefont {S.}~\bibnamefont
  {Pironio}},\ }\href {\doibase 10.1103/PhysRevA.68.062109} {\bibfield
  {journal} {\bibinfo  {journal} {Phys. Rev. A}\ }\textbf {\bibinfo {volume}
  {68}},\ \bibinfo {pages} {062109} (\bibinfo {year} {2003})}\BibitemShut
  {NoStop}%
\bibitem [{\citenamefont {Fine}(1982)}]{Fine-1982a}%
  \BibitemOpen
  \bibfield  {author} {\bibinfo {author} {\bibfnamefont {A.}~\bibnamefont
  {Fine}},\ }\href {\doibase 10.1063/1.525514} {\bibfield  {journal} {\bibinfo
  {journal} {Journal of Mathematical Physics}\ }\textbf {\bibinfo {volume}
  {23}},\ \bibinfo {pages} {1306} (\bibinfo {year} {1982})}\BibitemShut
  {NoStop}%
\bibitem [{\citenamefont {Bene}\ and\ \citenamefont
  {Vértesi}(2018)}]{Bene-2018a}%
  \BibitemOpen
  \bibfield  {author} {\bibinfo {author} {\bibfnamefont {E.}~\bibnamefont
  {Bene}}\ and\ \bibinfo {author} {\bibfnamefont {T.}~\bibnamefont
  {Vértesi}},\ }\href {\doibase 10.1088/1367-2630/aa9ca3} {\bibfield
  {journal} {\bibinfo  {journal} {New Journal of Physics}\ }\textbf {\bibinfo
  {volume} {20}},\ \bibinfo {pages} {013021} (\bibinfo {year}
  {2018})}\BibitemShut {NoStop}%
\bibitem [{\citenamefont {Hirsch}\ \emph {et~al.}(2018)\citenamefont {Hirsch},
  \citenamefont {Quintino},\ and\ \citenamefont {Brunner}}]{Hirsch-2018a}%
  \BibitemOpen
  \bibfield  {author} {\bibinfo {author} {\bibfnamefont {F.}~\bibnamefont
  {Hirsch}}, \bibinfo {author} {\bibfnamefont {M.~T.}\ \bibnamefont
  {Quintino}}, \ and\ \bibinfo {author} {\bibfnamefont {N.}~\bibnamefont
  {Brunner}},\ }\href {\doibase 10.1103/PhysRevA.97.012129} {\bibfield
  {journal} {\bibinfo  {journal} {Phys. Rev. A}\ }\textbf {\bibinfo {volume}
  {97}},\ \bibinfo {pages} {012129} (\bibinfo {year} {2018})}\BibitemShut
  {NoStop}%
\bibitem [{\citenamefont {Wolf}\ \emph {et~al.}(2009)\citenamefont {Wolf},
  \citenamefont {Perez-Garcia},\ and\ \citenamefont {Fernandez}}]{Wolf-2009a}%
  \BibitemOpen
  \bibfield  {author} {\bibinfo {author} {\bibfnamefont {M.~M.}\ \bibnamefont
  {Wolf}}, \bibinfo {author} {\bibfnamefont {D.}~\bibnamefont {Perez-Garcia}},
  \ and\ \bibinfo {author} {\bibfnamefont {C.}~\bibnamefont {Fernandez}},\
  }\href {\doibase 10.1103/PhysRevLett.103.230402} {\bibfield  {journal}
  {\bibinfo  {journal} {Phys. Rev. Lett.}\ }\textbf {\bibinfo {volume} {103}},\
  \bibinfo {pages} {230402} (\bibinfo {year} {2009})}\BibitemShut {NoStop}%
\end{thebibliography}%

\end{document}